\documentclass[twocolumn,trackchanges]{aastex63}
\usepackage[all]{hypcap}
\usepackage{float}

\graphicspath{{./}}

\begin{document}

\title{Evolution of elemental abundances in hot active region cores from Chandrayaan-2 XSM observations}

\correspondingauthor{Biswajit Mondal}
\email{biswajit70mondal94@gmail.com, biswajitm@prl.res.in}

\author[0000-0002-7020-2826]{Biswajit Mondal}
\affiliation{Physical Research Laboratory, Navrangpura, Ahmedabad, Gujarat-380 009, India }
\affiliation{Indian Institute of Technology Gandhinagar, Palaj, Gandhinagar, Gujarat-382 355, India}
\author[0000-0002-2050-0913]{Santosh V. Vadawale}
\affiliation{Physical Research Laboratory, Navrangpura, Ahmedabad, Gujarat-380 009, India }
\author[0000-0002-4125-0204]{Giulio Del Zanna}
\affiliation{DAMTP, Centre for Mathematical Sciences, University of Cambridge, Wilberforce Road, Cambridge CB3 0WA, UK}
\author[0000-0003-3431-6110]{N. P. S. Mithun}
\affiliation{Physical Research Laboratory, Navrangpura, Ahmedabad, Gujarat-380 009, India }
\author[0000-0002-4781-5798]{Aveek Sarkar}
\affiliation{Physical Research Laboratory, Navrangpura, Ahmedabad, Gujarat-380 009, India }
\author[0000-0002-6418-7914]{Helen E. Mason}
\affiliation{DAMTP, Centre for Mathematical Sciences, University of Cambridge, Wilberforce Road, Cambridge CB3 0WA, UK}
\author[0000-0003-2504-2576]{P. Janardhan}
\affiliation{Physical Research Laboratory, Navrangpura, Ahmedabad, Gujarat-380 009, India }
\author[0000-0003-1693-453X]{Anil Bhardwaj}
\affiliation{Physical Research Laboratory, Navrangpura, Ahmedabad, Gujarat-380 009, India }

\begin{abstract}

The First Ionization Potential (FIP) bias, whereby elemental abundances for low FIP elements in different coronal structures vary from their photospheric values and may also vary with time, has been widely studied. In order to study the temporal variation, and to understand the physical mechanisms giving rise to the FIP bias,
we have investigated the hot cores of three ARs using disk-integrated soft X-ray spectroscopic observation with the Solar X-ray Monitor (XSM) onboard Chandrayaan-2.
Observations for periods when only one AR was present on the solar disk were used so as to ensure that the AR was the principal contributor to the total X-ray intensity. The average values of temperature and EM were $\sim$3 MK and 3$\times10^{46}$ cm$^{-3}$ respectively. 
{ 
Regardless of the AR's age or activity, the elemental abundances for the low FIP elements Al, Mg, and Si with respect to the soft X-ray continuum were consistently higher than their photospheric values.
The average FIP bias for Mg and Si was 2 to 2.5,
whereas the FIP bias for the mid-FIP element, S, was almost unity.
However, the FIP bias for the lowest FIP element, Al, was observed to be a factor of 2 higher than Si,
which, if real, suggests a dependence of the FIP bias of low FIP elements on their FIP value.} 
Another major result from our analysis is that the
FIP bias of these elements is established within $\sim$10 hours of emergence of the AR and then remains almost constant throughout its lifetime.


\end{abstract}



\section{Introduction}
\label{Sec:Introduction}

The earlier study of the Sun as a star by \citet{pottasch_1963} revealed that solar 
coronal abundances are different from those of the photosphere.
The differences are correlated to the  First Ionization Potential (FIP) of the element,
in the sense that the abundance ratio of a low-FIP (less than 10 eV) element
versus that of a high-FIP element is higher in the corona. 
A measure of the difference is the so called FIP bias, i.e. the ratio
between the coronal and the photospheric abundance of an element.

{ The FIP bias is generally estimated by
measuring the relative abundances between elements, rather than to hydrogen. This is due to the fact that abundance measurements with respect to Hydrogen in the low corona are non-trivial. Hence, whether it is the low-FIP elements that have an increased abundance or the high-FIP elements that have a reduced one (compared to their photospheric values) is unclear.}



Further, it has become clear that different solar structures have different
FIP biases. There are also indications that the FIP bias depends on the temperature of the plasma. 
For a long time, it has been widely accepted that coronal abundances in active regions increase with time.
We refer the reader to the  recent reviews by \cite{Laming_2015,Zanna_2018LRSP} for more details.
We also provide, in the following section, a brief summary of available measurements related to active regions.

Knowledge of the elemental abundances in different atmospheric layers of the
Sun is a topic of great interest to the solar physics community mainly due to the following two reasons.
The first is that they provide, in principle, a way to link
the solar source regions to the various components of the solar wind. In fact, elemental abundance variations are  also clearly observed in-situ. The slow-speed solar wind has a high FIP bias similar to that measured in AR core loops, 3MK, whereas the high-speed wind has a near unit FIP bias, similar to that of coronal holes
(see, e.g., \citealp{Brooks_2015NatCo,Gloeckler_1989,Feldman_1998,Bochsler_2007,Brooks_2011ApJ}).

The second reason is that studying abundance variations contributes to a better understanding of the
physical processes at play in the solar corona. In fact, we know that the FIP bias  is closely related to the magnetic field activity of the Sun \citep[see, e.g.][]{Feldman_2002PhPl,Brooks_2017,Baker_2018ApJ}.
The Ponderomotive force model~\citep{Laming_2004,Laming_2009ApJ,Laming_2012ApJ,Laming_2017}
is now  widely accepted as being able to reproduce the main characteristics of the FIP effect,
as measured in-situ and remotely.
According to this model, the separation of ions from  neutral atoms within closed loops in an upward direction is caused by the reflection of downward propagating 
Alfv'en waves at chromospheric heights, causing
an enhancement of the low-FIP elements in the corona.
Since coronal waves can be produced by mechanisms that heat the solar corona, it is thought that the mechanism underlying the FIP effect is inextricably linked to processes that heat the solar corona. Hence, measuring the FIP bias is an important diagnostic for coronal plasma characteristics~\citep{Laming_2015, Dahlburg_2016}.

In this paper, we focus on the elemental abundances of hot, quiescent AR core emission at 3 MK, by providing line-to-continuum measurements of the Sun in the soft X-ray energy band using data from the Solar X-ray Monitor (XSM: \citealp{vadawale_2014,shanmugam_2020}). It may be noted here that the XSM is the only spectrometer to have observed the Sun in the 1-15 keV range during the minimum of solar cycle 24 with an energy resolution better than 180 eV at 5.9 keV.  This resolution is sufficient
to measure the abundances of several elements.



The XSM energy band is sensitive to temperatures above 2 MK.
When the Sun was at minimum activity levels, without any ARs, the XSM observed a steady signal
originating from X-ray Bright Points (XBPs), with a peak emission around 2 MK~\citep{xsm_XBP_abundance_2021}.
%
%
%
When a single non-flaring AR is present, the signal is dominated by the AR's near-isothermal
$\sim$ 3 MK emission \citep[see, e.g.][]{delzanna:2013}.
This provides an excellent opportunity to measure the FIP bias or abundance of the 
hot AR core with respect to the continuum.
{ 
However, the soft X-ray continuum for 3 MK AR plasma
is dominated by free-bound
emission, which depends on abundant elements (e.g., O, Ne, etc), as detailed in the Appendix. Therefore, the measured abundances are actually relative abundances.}

In the literature, few abundance measurements are known to be associated specifically
with the 3 MK emission from quiescent AR cores. These are summarised in
\cite{Zanna_2018LRSP}. X-ray spectra in the 10--20~\AA\ range have provided
the relative abundances of the low-FIP Fe, Mg vs. O, Ne.
Most studies provided results on single active regions. 
\cite{Saba_1993AdSpR} reported a significant variability of the FIP bias using  SMM/FCS observations
of several active regions.
On the other hand, a re-analysis of several quiescent AR cores with improved atomic data
and using a multi-thermal DEM technique by \cite{Zanna_2014_AR_abund} indicated the same
FIP bias, around 3, for all the active regions studied, irrespective of their age and size.

Since 2006, EUV spectra from the Hinode EIS instrument have provided an opportunity to
measure the relative FIP bias between low-FIP elements (e.g. Fe, Si) and the high-FIP, Ar,
as well as the mid-FIP, S. The latter actually shows the same abundance variations as the
high-FIP elements.  An example case was discussed by \cite{delzanna:2013},  showing that the FIP bias
in the EUV of 3 MK plasma was the same as in the X-rays.
Considering the size of the emitting plasma and its emission measure,
\cite{delzanna:2013} concluded that it should be the low-FIP elements that
are over-abundant by about a factor of 3. 

\cite{Zanna_2022ApJ} carried out a multi-wavelength study of an AR as it crossed the
solar disk which 
was observed by XSM as well as by SDO/AIA, Hinode/EIS and Hinode/XRT.
The relative FIP bias obtained from Hinode/EIS observations confirmed the \cite{delzanna:2013} results,
and showed no variation with the disk passage.  
{ The  analysis of simultaneous XSM spectra on two days
also indicated no significant variability, and provided an FIP bias for Si of 2.4}, i.e. close to the value suggested by \cite{delzanna:2013},
and also very close to the prediction of Laming's model.

In the present study, we extend the previous XSM analysis to all the quiescent
periods of that same active region, and also investigate two other active regions
during their disk crossings. One AR in particular is of interest as it emerged on-disk, and
hence offers the opportunity to study the elemental abundances during the early
phase of the evolution of an AR.

The rest of the paper is organized as follows: Section~\ref{sec:overview} provides a
short overview of previous abundance measurements in active regions.
Section~\ref{sec:obs} describes the observations and data analysis. Section~\ref{sec:SpectralAnalysis} provides a detailed spectral analysis. After obtaining the results, these are discussed in Section~\ref{sec:results_discussion}. Section~\ref{sec:summary} provides a brief summary of the article.

\section{Historical overview}
\label{sec:overview}


Spatially resolved measurements of the relative FIP bias have been carried out by several
authors 
\citep[see,e.g.][]{widing_feldman:1993,Sheeley_1995ApJ,Sheeley_1996ApJ,Widing_1997,Widing_2001ApJ} using 
Skylab spectroheliograms with Mg, Ne transition region lines. These are formed well below 1 MK,
in the legs of `cool' (1 MK) AR loops. They found 
photospheric composition (FIP bias=1) for  newly emerged closed loops, but increasing
FIP bias reaching a value of 3-4 within a timescale of 1-2 days~\citep{Widing_2001ApJ},
and much higher values, up to about 10, within a few more days.
 Differing FIP biases were also obtained by \cite{Young_1997SoPh} and \cite{Dwivedi_1999ApJ} using  Mg 
and Ne line ratios observed by the
CDS and SUMER spectrometers onboard the Solar and Heliospheric Observatory (SOHO). 

The large values for the FIP bias (around 10) are hard to reconcile with in-situ measurements, where the
FIP bias is at most 3, and also with theory.
However,  \cite{zanna_2003A&A} pointed out that as the cool AR loops are almost
isothermal in their cross-section, the assumption that a smooth emission measure
distribution was present in the plasma, used to interpret the Skylab data, was not justified.
\cite{zanna_2003A&A} took the  intensities measured by \cite{widing_feldman:1993},
and using an emission measure
loci approach, showed that a FIP bias of 3.7 was consistent with the data, much
lower than the value of 14 reported by Widing and Feldman. 
\cite{zanna_2003A&A} also analysed the legs of several cool loops observed by
SoHO/CDS and found photospheric abundances, although a similar analysis for
other loops by \cite{zanna_2003A&Ab} found a FIP bias of 4.

In summary, the legs of cool AR loops do show a range of FIP bias values,
between 1 and 4, and perhaps occasionally larger.
However, the very high FIP biases found from Skylab data were
largely overestimated.  


As shown by \cite{zanna_2003A&Ab}, active region cores are composed
not only of cool 1 MK loops and unresolved, almost isothermal 3 MK loops, but also
unresolved emission in the 1--3 MK range. The plasma at different temperatures
is generally not co-spatial. 

There is evidence from Hinode EIS observations of
e.g. Si X, S X lines that this $\simeq$2 MK emission has a lower relative
FIP bias, around 2
\citep[see,e.g.][]{Zanna_2012A&A}.
Further studies using the same lines
(e.g., \citealp{Baker_2013ApJ,Baker_2015ApJ,Doschek_2019ApJ,Mihailescu_2022ApJ,Ko_2016ApJ,Testa_2022arXiv})
have shown some variation (around the value of 2) of the
relative FIP bias within each active region, but little variability in time,
except during the decay phase, when an AR effectively disappears and
the relative abundances become photospheric.

In summary, active region structures formed at temperatures below 2 MK
show a range of relative FIP biases, and some temporal variability.
The few observations of the hotter, 3 MK, AR cores have in contrast shown a remarkable
consistency, with relative FIP biases around 3.

Finally, to interpret observations of the Sun as a star, one needs
to take into account the above (and other) issues.
As shown by \cite{delzanna:2019_eve}, when the Sun's actvity is at a minimum with no active region present on the solar disk,
the corona around 1 MK shows near photospheric abundances, 
whereas in presence of active regions, the FIP bias for the 1 MK emission stays
the same, but the hotter emission shows a higher relative FIP bias.
When active regions flare, the high temperature plasma shows
nearly photospheric composition around the peak X-ray emission \citep[see e.g.,][]{Mondal_2021ApJ}.




\section{Observations and data analysis}
\label{sec:obs}

Observations of the Sun were carried out with the XSM during the minimum of solar cycle 24, when no active regions were present, covering the years 2019-2020. Results are given in~\cite{xsm_XBP_abundance_2021}. 
They reported intermediate abundances of low-FIP elements (Mg, Al, and Si) of 2 MK plasma, primarily originating from X-ray Bright Points, XBPs~\citep{Mondal_2022ApJ}. 
Frequent  micro-flaring activity was observed and found to be occurring everywhere on the solar disk, even when no ARs were present~\citep{xsm_microflares_2021}.
%
%
During the minimum of solar cycle 24, XSM observed the disk passage of a few individual, isolated ARs in the absence of any other major activity. 
When ARs were present on-disk, XSM recorded hundreds of small flares of different classes. Elemental abundance variations during these small flares were found, for the first time, to initially drop towards photospheric values, then rapidly return to coronal values, as described by \cite{Mondal_2021ApJ}, \cite{Mithun_2022ApJ}, and  \cite{laksitha_2022}.
In this paper, we analyze the temporal 
evolution of active regions outside of flaring activity and for this we have 
chosen to study  three isolated
active regions: AR12749, AR12758, and AR12759.

XSM data contain spectra at 1 s cadence in a raw (level-1) daily file. Since the visibility of the Sun varies within the XSM field-of-view (FOV), with the Sun being sometimes outside the FOV or being occulted by the Moon, the data include both solar and non-solar spectra.
The XSM Data Analysis Software (XSMDAS:~\cite{xsm_data_processing_2020}) has been used to generate the level-2 science data product using the appropriate Good Time Intervals (GTIs) and the other necessary instrumental parameters.
The available default level-2 data contains the effective area corrected light curves for every second and spectra for every minute. 
XSMDAS also provides the functionality to generate the light curves and spectra for a given cadence and energy range, which we have used in the present analysis.

\begin{figure*}[!ht]
\begin{center}
    \includegraphics[width=0.8\textwidth]{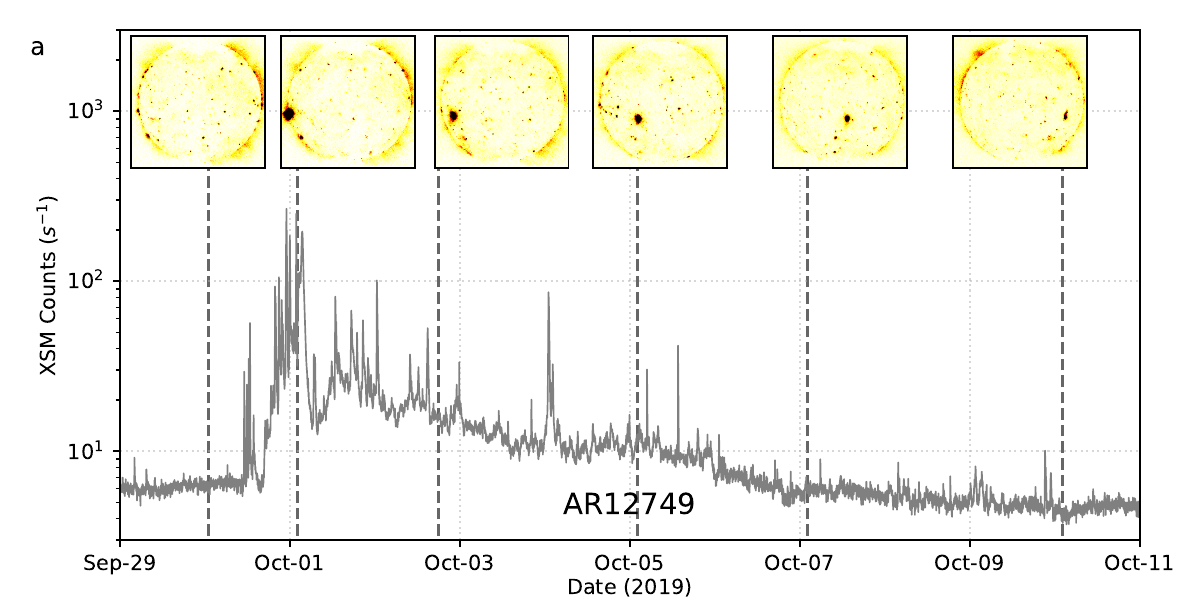}
    \includegraphics[width=0.8\textwidth]{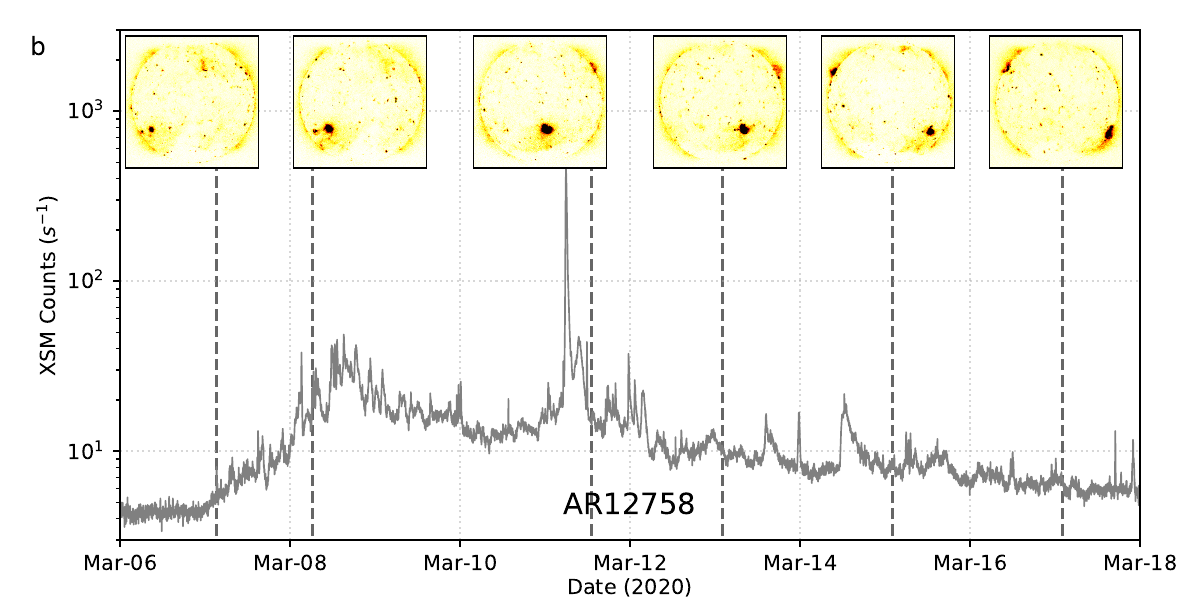}
    \includegraphics[width=0.8\textwidth]{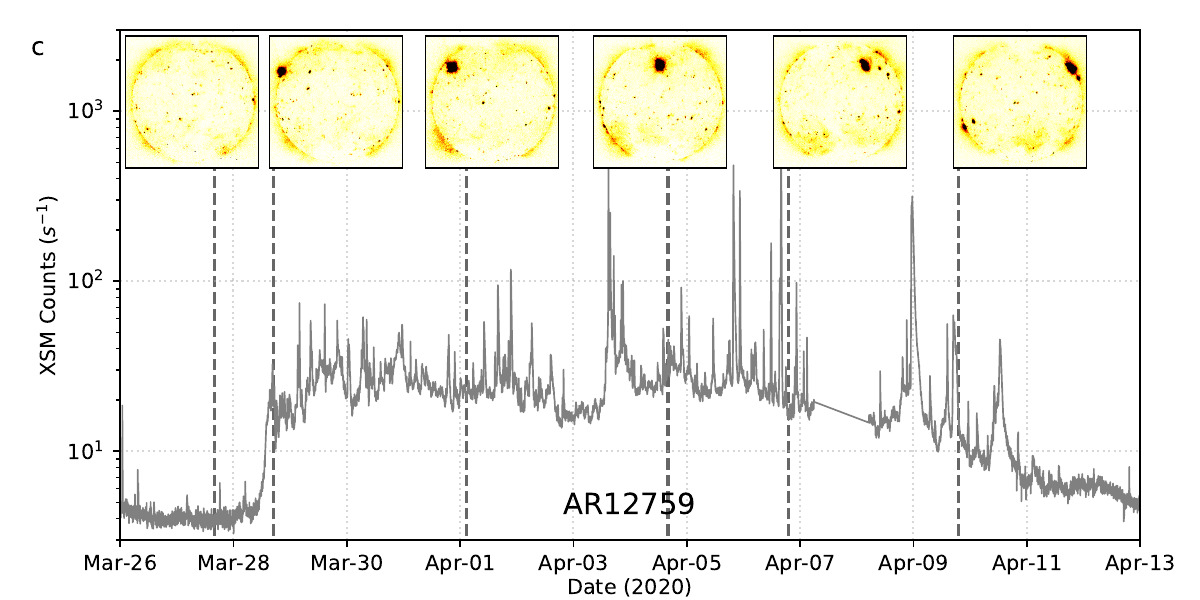}
    \caption{XSM 1-15 keV light curves during the disk passage of AR12749 (panel a), AR12758 (panel b) and AR12759 (panel c).
    The top row of each panel shows   representative full disk X-ray images (negative intensities) taken with the XRT Be-thin filter during the evolution of the ARs. The vertical dashed lines represent the timing of the XRT images.
    	}
    \label{fig:ARs_LC}
\end{center}
\end{figure*}

 Using the XSMDAS, we have generated  2 min averaged XSM light curves in the energy range of 1-15 keV during the disk passage of the AR12749, AR12758, and AR12759, as shown in the three panels of Figure~\ref{fig:ARs_LC}.
 During the evolution of these three ARs, representative full disk X-ray images taken by the XRT Be-thin filter are shown in the top row of each panel. 
  AR12749 (Figure~\ref{fig:ARs_LC}a) appeared from the east limb on Sept 29, 2019.
  Whilst crossing the solar disk, it became fainter towards the west limb and went behind the limb on 14 Oct. 
  AR12758 (Figure~\ref{fig:ARs_LC}b) appears to form on disk on 06 Mar 2020 and has fully emerged after 08 Mar.
  It decays whilst crossing the solar disk and finally goes behind the west limb on 18 Mar. 
  AR12759 appeared from the east limb on 29 Mar 2020 and transited the solar disk until 14 Apr 2020, before disappearing behind the west limb.  
  
 The full disk XRT images show that during the passage of these three ARs, no other major activity was present on the solar disk. Thus, we conclude that these three ARs were primarily responsible, during their disk passage, for the enhanced X-ray emission observed by the XSM.
 These ARs produced many small B/A-class flares, seen as multiple spikes in the XSM light curves. Detailed studies of these small flares were reported by \cite{Mondal_2021ApJ} and \cite{laksitha_2022}. 
 
\begin{figure}[!ht]
\begin{center}
    \includegraphics[width=0.55\textwidth]{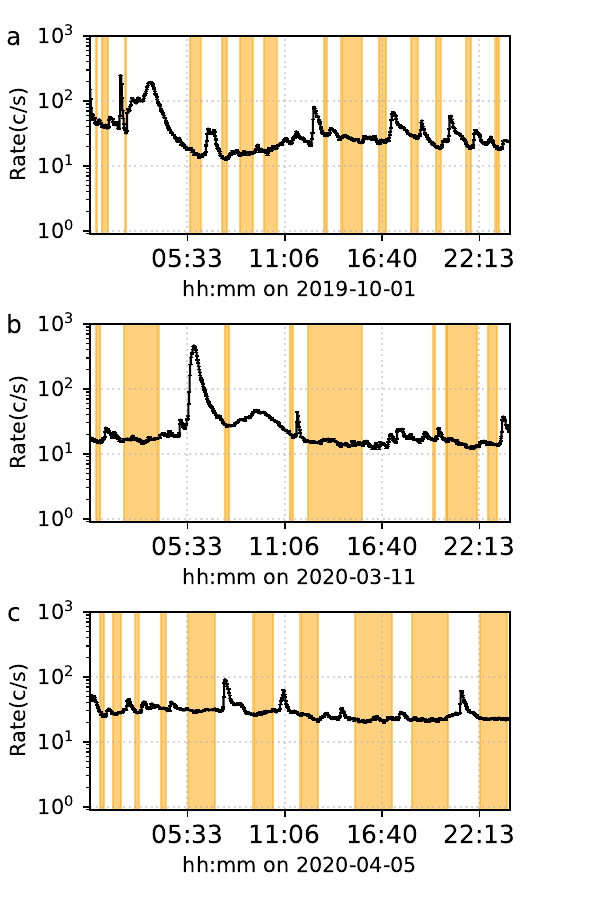}
    \caption{Selection of the quiescent AR periods (orange-shaded regions) from the XSM light-curves for one representative day of AR12749 (panel a), AR12758 (panel b), and AR12759 (panel c).}
    \label{fig:LC_20200406}
\end{center}
\end{figure}

In the present study, 
we have selected only the quiescent periods from the observed light curves by excluding the periods when the small flares occurred using a semi-automated graphical algorithm. 
{ For example, Figure~\ref{fig:LC_20200406} shows the representative selection  (orange shaded regions) for one day of AR12749 (panel a), AR12758 (panel b), and
AR12759 (panel c).}
These identified time intervals were used as user-defined GTIs to generate the spectra for quiescent ARs on a daily basis in order to carry out the detailed spectral analysis as discussed in Section~\ref{sec:SpectralAnalysis}.

\section{Spectral analysis}\label{sec:SpectralAnalysis}

Broad-band soft X-ray spectra of the solar corona consist of a continuum as well as the emission lines of the different elements. Modeling the soft X-ray spectrum provides the measurements of the temperature, emission measure, and elemental abundances {(with respect to continuum emission)} of the emitting plasma~\citep{Zanna_2018LRSP}.
We use the \verb|chisoth| model~(see Appendix of, \citealp{Mondal_2021ApJ}) for the spectral fitting.
The \verb|chisoth| is a local model of the X-ray spectral fitting package (XSPEC:~\cite{ref-xspec}), and it estimates the theoretical spectrum  using the CHIANTI atomic database.
It takes temperature, emission measure (EM: which is related to the density of the plasma), and the elemental abundances of the elements from Z=2 to Z=30 as free variables for the spectral fitting.

After generating the spectra for the quiescent periods, we fitted them with an isothermal emission model.
For the spectral fitting, we ignored the spectra below 1.3 keV where the XSM response is not well-known ~\citep{xsm_flight_performance}, 
and above the energy where the solar spectrum is dominated by the non-solar background spectrum.
During the spectral fitting, the temperature, EM, along with the abundances of Mg, Al, and Si (whose emission lines are prominent in the XSM  spectrum) were kept as variable parameters. The 1$\sigma$ uncertainty of each free parameter was also estimated using the standard procedure in XSPEC.

Although the S line complex is visible in the spectra, including it in the spectral fits as a free parameter causes a large uncertainty in the measurement of the S abundance because of its poor statistics.
Hence, in \verb|chisoth| model, we fixed the S abundances along with the abundances of other elements (whose emission lines are not visible in the observed spectra) with the photospheric abundances of \cite{Asplund_2009}. 
{ However, we found that the measurement of the S abundance is possible for the statistically improved summed spectra of the entire AR period.
}


{ It should be noted that the measured abundances are with respect to the continuum emission, which has contributions from free-free, two-photon, and free-bound processes (see Appendix~\ref{app_emission_procc}). 
For AR temperatures of around 3 MK, the main contribution to the continuum is free-bound,
mostly (nearly
80\%) by O and Ne (see Appendix~\ref{app_emission_procc}).
As the observed spectra do not have any emission lines of O and Ne, their abundances can not be measured from the fitting.
Therefore,  in our abundance measurements, we have to assume a reasonable value of the O and Ne abundances for an active region. 
In measurements and models of the FIP effect, it is common to use Oxygen as a reference point, and measure other abundances relative to Oxygen.  As the FIP for Oxygen is close to that of Hydrogen, 
it is often assumed that it is not fractionated, hence has a photospheric abundance.

A direct measurement of O/H abundance is challenging.
However, we do have direct in-situ measurements of O/H abundance in the solar wind (SW).
The measurements of the O/H abundance in the SW show
a scatter centered around 8.8 dex (cf. \citealp{Bame_1975SoPh, von_Steiger_2010}), which is close to the  photospheric value ($\sim$8.7 dex) recommended by~\citep{Asplund_2009,Asplund_2021A&A}.
As most  high-FIP elements such as Ne  show similar abundance variations as Oxygen, it is reasonable to assume that coronal abundance of Ne ( C, N) also will have a value close to photospheric.
Therefore, we have fixed the abundances of Ne, C, and N to the 
values recommended by~\cite{Asplund_2009}.
{It is noted that fixing the abundance of these elements, except O, to their coronal values reported in the literature (e.g., \citealp{Feldman_1992, Fludra_1999}) does not vary the measured abundance of Mg, Al, Si, and S significantly. They remain within the error bars.}

}

\begin{figure}[!ht]
\begin{center}
    \includegraphics[width=0.5\textwidth]{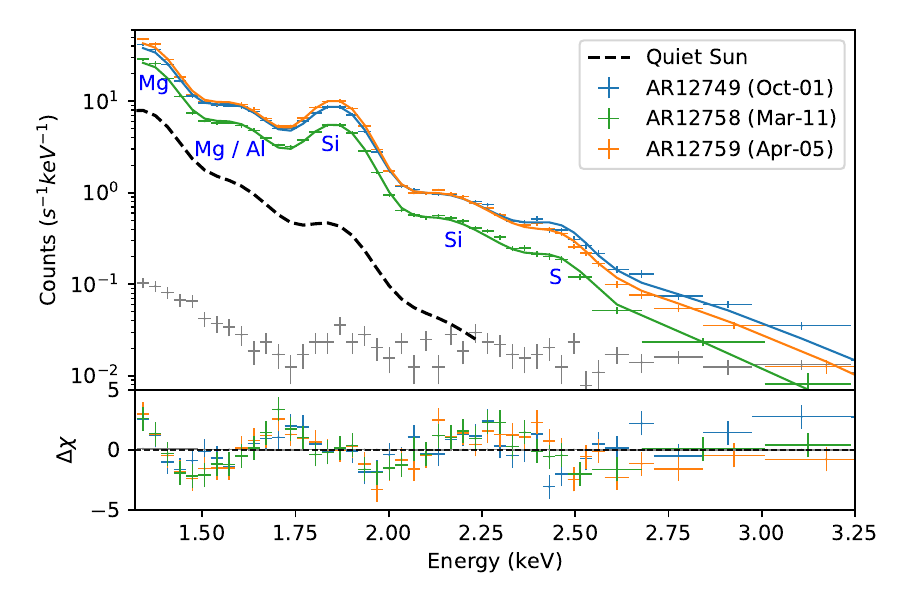}
    \caption{Soft X-ray spectra measured by the XSM for three representative days of the AR period are shown. Solid lines represent the best-fit isothermal model, and the residuals are shown in the bottom panel. Gray points correspond to the non-solar background spectrum. 
\label{fig:AR_1T_fitSpec}}
\end{center}
\end{figure}

Figure~\ref{fig:AR_1T_fitSpec} shows the representative XSM spectra, for the three ARs fitted, in different colours, with an isothermal model. 
The points with error bars represent the observed spectra, whereas the solid curves represent the best-fit modeled spectra. The grey error bars represent the non-solar background spectrum, which is subtracted from the observed spectra during the spectral analysis. The lower panel shows the residual between the observed and model spectra. 
We have fitted all the spectra in a similar way and found that all of them are well described by an isothermal model.

The X-rays observed by XSM originated from both the AR and the background quiet Sun regions (outside the AR).
To determine how much emission is due to the background quiet Sun regions, we estimate the average quiet Sun spectrum using an average quiet Sun temperature, EM, and abundances, as reported by \cite{xsm_XBP_abundance_2021}. 
The average quiet Sun spectrum is shown by the black dashed curve in Figure~\ref{fig:AR_1T_fitSpec}.
The quiet Sun spectrum is found to be almost an order of magnitude lower than the spectra of the active period 
when the ARs were very bright on the solar disk.
We thus conclude that the X-ray emission of the active periods is primarily dominated by the AR emission.

Separating the AR emission from the background quiet Sun emission would be possible by subtracting the quiet-sun spectra from the AR spectra. But, as the effective area of the XSM varies with time, this is not recommended. It is possible to model the AR spectra using a two-temperature (2T) component model rather than subtracting the quiet Sun spectra. This is what we have chosen to do.
%
%
One temperature corresponds to the background solar emission originating from the regions outside the AR and the second temperature corresponds to the AR plasma.
We have modeled a few AR spectra with a two-temperature (2T) model.
During the 2T spectral fitting, the parameters of the background solar emission were kept fixed to the average quiet Sun values reported by \cite{xsm_XBP_abundance_2021}. 
For the AR component, the temperature, EM, along with the abundances of Mg, Al, and Si, were kept as variable parameters.
We found that the 2T model can describe the XSM spectra for the active periods with similar best-fitted parameters as those obtained by the isothermal model.
This verifies that the AR emission dominates the spectra of the AR periods. 
Thus, in this study, we show the  results of the isothermal analysis in Figure~\ref{fig:AR_T_EM} and \ref{fig:AR_abund}. This is discussed in Section~\ref{sec:results_discussion}.

\begin{figure*}[!ht]
\begin{center}
    \includegraphics[width=0.99\textwidth]{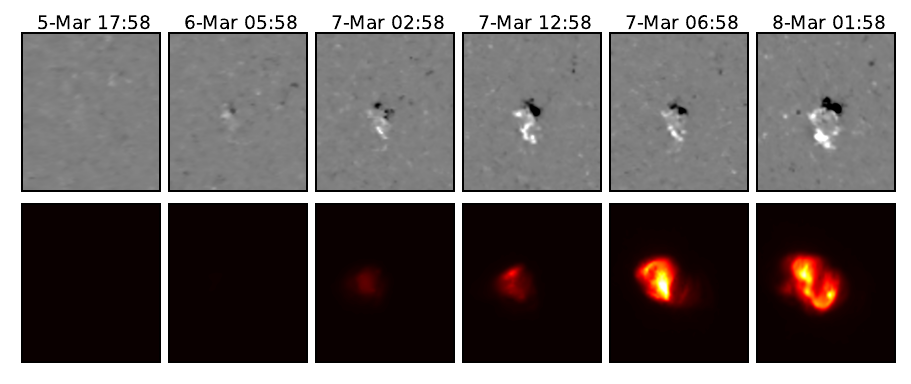}
    \caption{Evolution of the AR12758 during its emergence phase on the solar disk. Top row shows the evolution of photospheric magnetograms as observed by HMI and bottom row shows the evolution of X-ray emission as observed by XRT Be-thin filter.}
\label{fig:HMI_images_AR12758}
\end{center}
\end{figure*}

It is interesting to study how the plasma parameters 
vary during the emerging phase of the AR12758, i.e., from 07-Mar-2020 to 09-Mar-2020.
Figure~\ref{fig:HMI_images_AR12758} shows the evolution of the photospheric magnetograms (top row) and the X-ray emission (bottom row) as observed by SDO/HMI
and the Be-thin filter of Hinode/XRT respectively.
These images were created by de-rotating the synoptic data of HMI\footnote{http://jsoc.stanford.edu/data/hmi/synoptic/} and XRT\footnote{http://solar.physics.montana.edu/HINODE/XRT/SCIA/} 
to a common date (08-Mar-2020) using the standard procedure of SolarSoftWare (SSW; \citealp{Freeland_1998}).
We also determined the total unsigned photospheric magnetic flux for the regions $\pm$10 G within the field-of-view shown in Figure~\ref{fig:HMI_images_AR12758}.
During this emerging flux period, we carried out a time-resolved spectroscopic study using the XSM observations with finer time bins of less than a day.
However, during this period, as the emission from the AR was not very bright, the emission from the AR and the rest of the Sun could have been mixed together. Thus to derive the evolution of the plasma parameters during this period, we modeled the observed XSM spectra with a 2T model, where one component represents the emission from the AR, and the other represents the emission from the rest of the Sun, as discussed in the previous paragraph. The results are shown in Figure~\ref{fig:FitParAR12758_emarge} and discussed in Section~\ref{sec:results_discussion}.

\section{Results and Discussion} 
      \label{sec:results_discussion}      

%
%

\begin{figure*}[!ht]
\begin{center}
    \includegraphics[width=1\textwidth]{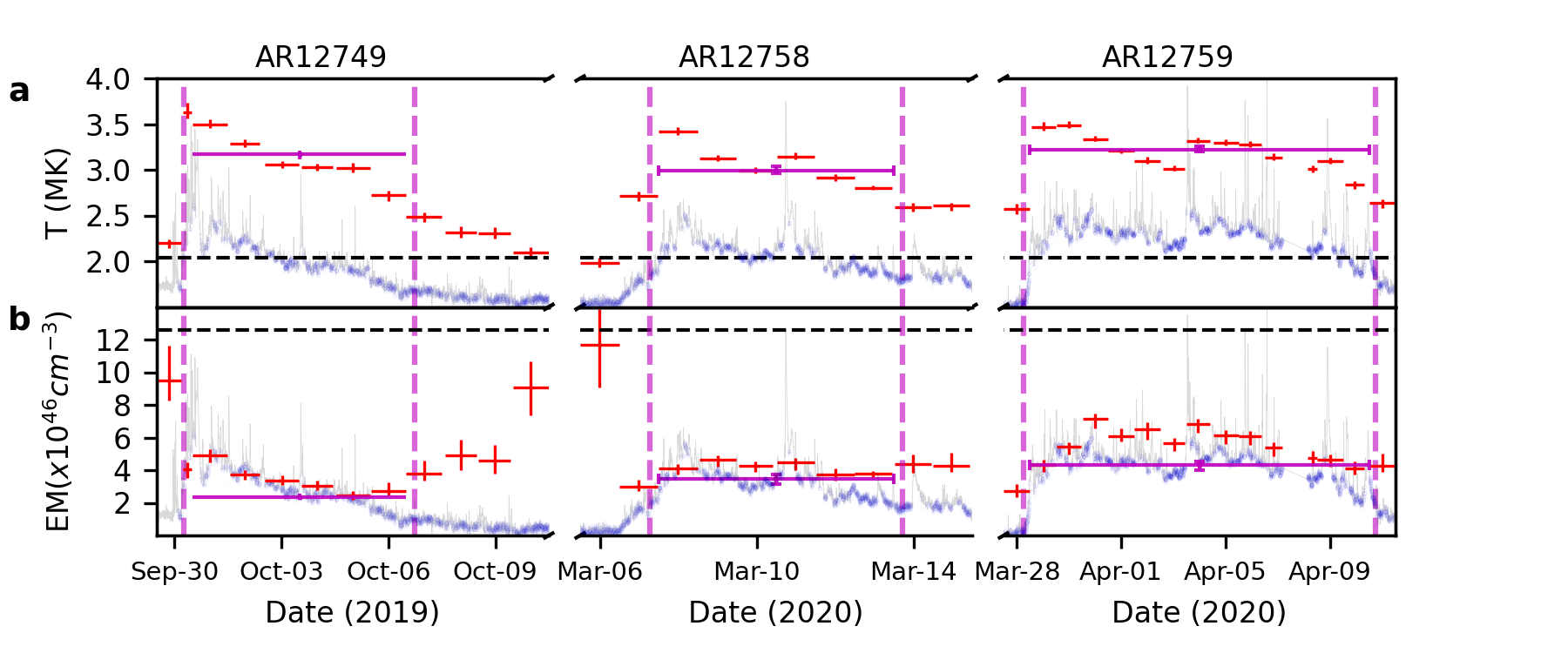}
    \caption{Evolution of the temperature (red points in panel a) and EM (red points in panel b) during the evolution of AR12749, AR12758, and AR12759. 
    When the ARs were very bright, as bounded by the vertical dashed lines, the magenta bars represent the average values of the temperature and EM.
    The black horizontal dashed lines represent the average temperature and emission measure for the quiet Sun in the absence of any AR reported by \cite{xsm_XBP_abundance_2021}.
    The XSM lightcurves of the ARs are shown in grey color, and the lightcurves for the quiescent regions are shown in blue colors. 
\label{fig:AR_T_EM}}
\end{center}
\end{figure*}

\begin{figure*}[!ht]
\begin{center}
    \includegraphics[width=1\textwidth]{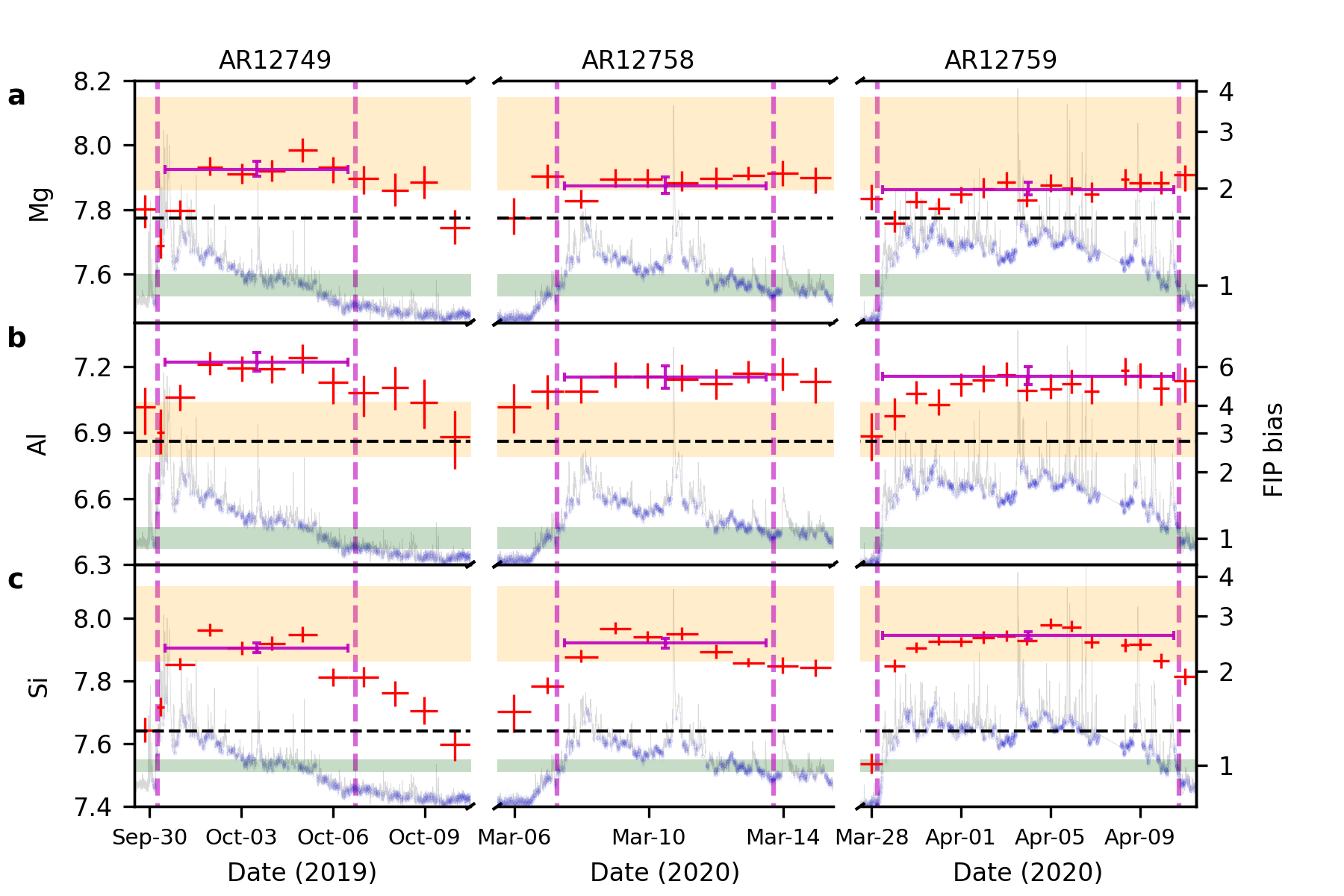}
    \caption{
    Panels a-c (red error bars) show the evolution of abundance in the logarithmic scale with A(H)=12 for Mg, Al, and Si during the evolution of AR12749, AR12758 and AR12759. 
    {  These abundances are measured with respect to the soft X-ray continuum, which is mostly determined by the choice of Oxygen abundance, which was considered to be 8.8 dex.}
    The magenta bars represented the average abundances when the ARs were very bright, as bounded by the vertical dashed lines. 
    The y-error bars represent 1$\sigma$ uncertainty for each parameter, and the x-error bars represent the duration over which a given spectrum is integrated. 
    The black horizontal dashed lines represent the average abundances for the quiet Sun in the absence of any AR reported by \cite{xsm_XBP_abundance_2021}.
    XSM light curves for each AR are shown in gray in the background, and the blue color on the XSM light curves represents the time duration excluding the flaring activities. 
    The range of coronal and photospheric abundances from various authors compiled in the CHIANTI database are shown as orange and green bands. The right y-axis shows the FIP bias values for the respective elements with respect to average photospheric abundances. 
\label{fig:AR_abund}}
\end{center}
\end{figure*}

In this study, we have performed the X-ray spectral analysis for the evolution of three ARs as observed by the XSM.
The AR spectra (Figure~\ref{fig:AR_1T_fitSpec}) show a clear signature of the thermal X-ray emission from the line complexes of Mg, Al, Si, and S, along with the continuum emission up to $\sim$3.0 keV.
The red points in Figure~\ref{fig:AR_T_EM} show the evolution of the temperature and EM throughout the evolution of the three ARs.
Figure~\ref{fig:AR_abund} shows the evolution of abundances  of Mg (panel a), Al (panel b), and Si (panel c).
The error bars associated with all the parameters along the y-axis represent the 1$\sigma$ uncertainties.
We also derived the average S abundance along with the other elements from the summed spectrum for the duration when the ARs were very bright on the solar disk (bounded by the vertical dashed lines in Figures~\ref{fig:AR_T_EM} and \ref{fig:AR_abund}). This provides the average parameters associated with each AR, as  shown by magenta bars and also given in Table~\ref{table-1}. 
%
%
The primary findings of the paper are discussed below.

\subsection{Temperature and emission measure}

Temperatures (T) and emission measures (EM) are close to the quiet Sun levels (black dashed lines in Figure~\ref{fig:AR_T_EM}) when the ARs were absent from the solar disc or only partially present, e.g., 30 September 2019 and 6 March 2020.
Once the ARs appear, the temperature rises to more than $\sim$3 MK from the $\sim$2 MK of the quiet Sun.
As the $\sim$3 MK emission is predominantly derived from a smaller volume of AR plasma, the presence of the AR reduces the EM from the quiet Sun values.
The average temperatures for all the ARs are determined to be $\sim$3 MK (magenta error bars in Figure~\ref{fig:AR_T_EM}a), which is close to the ``basal" temperature of the AR core reported in earlier research (e.g.,~\citealp{Zanna_2018LRSP,Zanna_2012A&A,Winebarger_2012ApJ}).
The temperature and EM do, however, vary slightly over the course of the AR's evolution, which is consistent with the observed X-ray light curve.
Following the arrival of AR12749 and AR12758, their activity decayed while rotating on the solar disk (Figure~\ref{fig:ARs_LC}), which is why the temperature and EM decreased during their evolution, as indicated by the dashed vertical lines in Figure~\ref{fig:AR_T_EM}.
After October 6, 2019, the EM for AR12749 begins to rise as the AR weakens and the quiet Sun emission takes precedence over the AR emission.
Thus, after the AR has almost died and is very faint, the EM and temperature reach values close to the quiet Sun temperature and EM.
The temperature and EM for the AR12759 remain almost constant with time, as this AR  crossed the solar disk without much decay in activity (Figure~\ref{fig:ARs_LC}c). 

\subsection{Abundance evolution}\label{discu:abund_evol}
In contrast to the temperature and EM, the abundances of Mg, Al, and Si do not follow the X-ray light curve of any of the three ARs throughout their evolution (Figure~\ref{fig:AR_abund}). 
The abundances obtained for low-FIP elements Al, Mg, and Si are consistently greater than the photospheric values, demonstrating a persistent FIP bias during the course of the AR.
{ 
Note that the abundances are measured with respect to the continuum emission, which depends on the abundances of other elements, primarily oxygen. 
}

After the emergence of AR12758, the FIP bias was found to be almost constant throughout its decay phase. 
Similarly, during the decay of the AR12749, the FIP bias remains nearly constant, in contrast to certain earlier studies, such as \cite{Ko_2016ApJ}. They suggested a decreasing FIP bias in high-temperature plasma of more than two million degrees during the decay phase of an AR. 
The more established AR, AR12759, which evolved without decaying much during its transit across the solar disk, also shows an almost constant FIP bias, similar to the other two ARs.

We do not find any relationship between the age of the AR and the FIP bias, as suggested in some previous papers, e.g.,\citealp{Zanna_2014_AR_abund, Doschek_2019ApJ}. 
{ Taking the oxygen abundance as 8.8 dex, the measured abundances for Mg and Si are within the range of coronal abundances reported in the literature (e.g.,  \citealp{Feldman_1992,Fludra_1999}, orange shaded regions in Figure~\ref{fig:AR_abund}).
However, the Al abundance is $\sim$30$\%$-60$\%$ higher.
}
We note that the Al lines in the XSM spectra are blended with Mg lines. From Markov Chain Monte Carlo (MCMC) analysis (discussed in Appendix A), we find that there is no anti-correlation between Mg and Al abundances. This suggests that the observed spectra does indeed require higher abundances of Al and cannot be explained by an enhancement of Mg abundances.

\subsection{FIP bias during the emergence of the AR core}
\begin{figure}[!ht]
\begin{center}
    \includegraphics[width=0.55\textwidth]{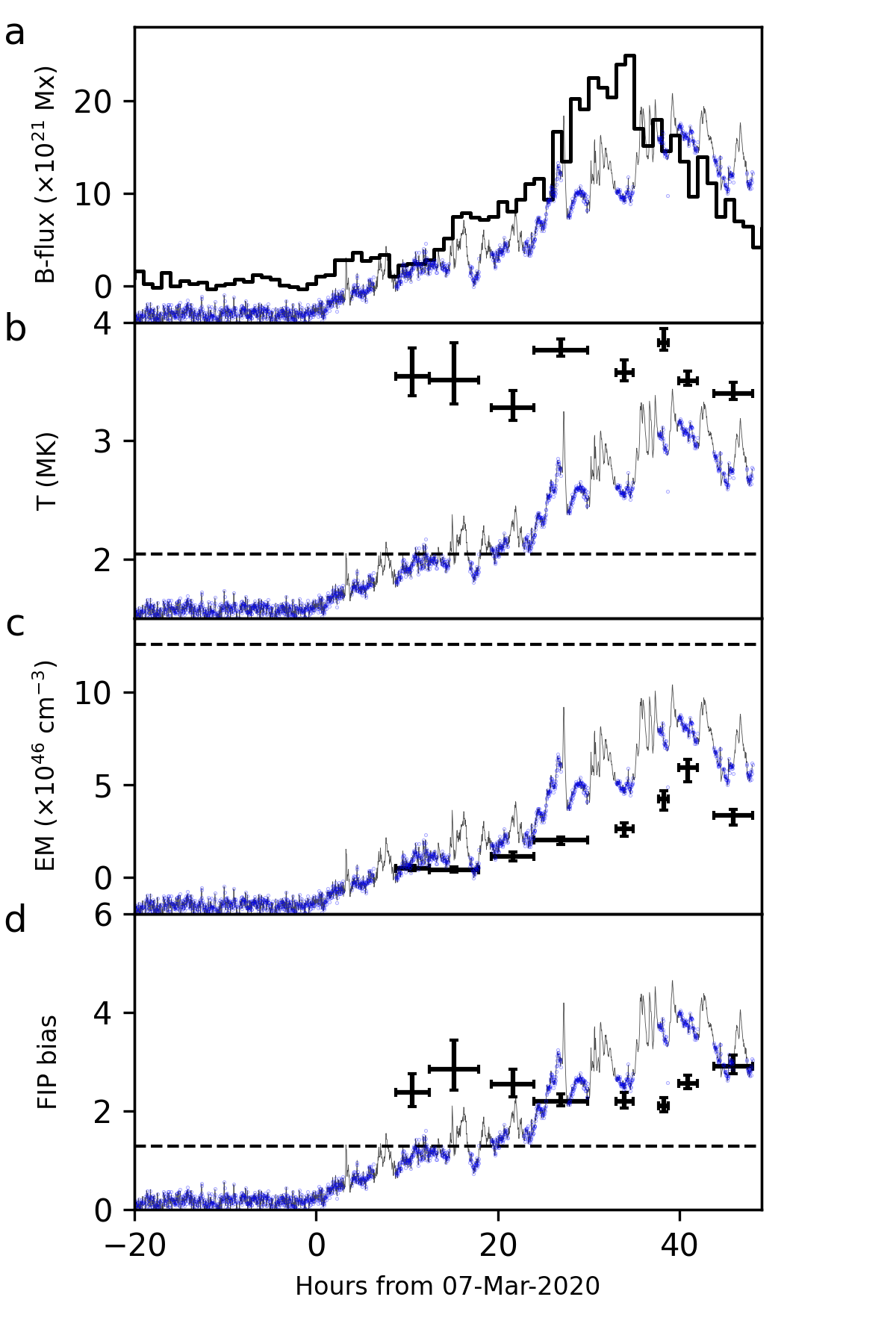}
    \caption{Results showing the emerging phase of AR12758. The black curve in panel a shows the evolution of the total unsigned photospheric magnetic flux. Panel b and c show the evolution of temperature and EM. Panel d shows the evolution of FIP bias for Si. The dashed lines in panels b-d represent the corresponding parameter for the background solar emission from the rest of the solar-disk except AR. The background grey curves in each panel represent the X-ray light curve observed by XSM. Whereas the blue curves represent the selected times excluding the flaring period, representing the quiescent AR. 
\label{fig:FitParAR12758_emarge}}
\end{center}
\end{figure}

Though we do not find any relationship between the age of the AR cores and their FIP biases (Section~\ref{discu:abund_evol}), which remain constant, it is interesting to study the timescale on which the FIP bias developed during the emergence of the AR core. 
Such a study has been made possible using the finer (less than a day) time-resolved spectroscopy during the emerging phase (07-Mar-2020 to 09-Mar-2020) of AR12758.
During this period, we estimated the total unsigned photospheric magnetic flux as measured by HMI/SDO and shown in Figure~\ref{fig:FitParAR12758_emarge}a (black color). 
The peak in the magnetic flux represents the time when the AR completely emerged onto the solar disk. 

After the emergence, the unsigned magnetic flux is found to (temporarily) decrease.
Figures~\ref{fig:FitParAR12758_emarge}b and \ref{fig:FitParAR12758_emarge}c show the evolution of the AR core temperature and emission-measure. With the emergence of the AR. The temperature becomes close to the AR core temperature of $\sim$3 MK, and the EM increases as the emitting plasma volume increases until it has emerged completely.
We also derived the evolution of the FIP bias during this period, shown in  Figure~\ref{fig:FitParAR12758_emarge}d for Si. During this period, as the emission from the Mg and Al line complex was weak compared with the background solar emission, the derived FIP bias for Mg and Al has a large uncertainty and is not shown here.
%
{  
Within $\sim$10 hours of the AR emergence,  the FIP bias was already close to 2.5, and remained almost constant throughout the evolution. }
So the emerging hot core loops for this AR do not show any variation in the FIP bias,
in agreement with previous suggestions. 
Recall that the variations in FIP bias reported earlier (e.g., \citealp{Widing_2001ApJ}) were observed in the cool loops, not the hot core loops. 

\subsection{Enhanced bias for Al}
\begin{figure}
\centering
    \includegraphics[width=0.5\textwidth]{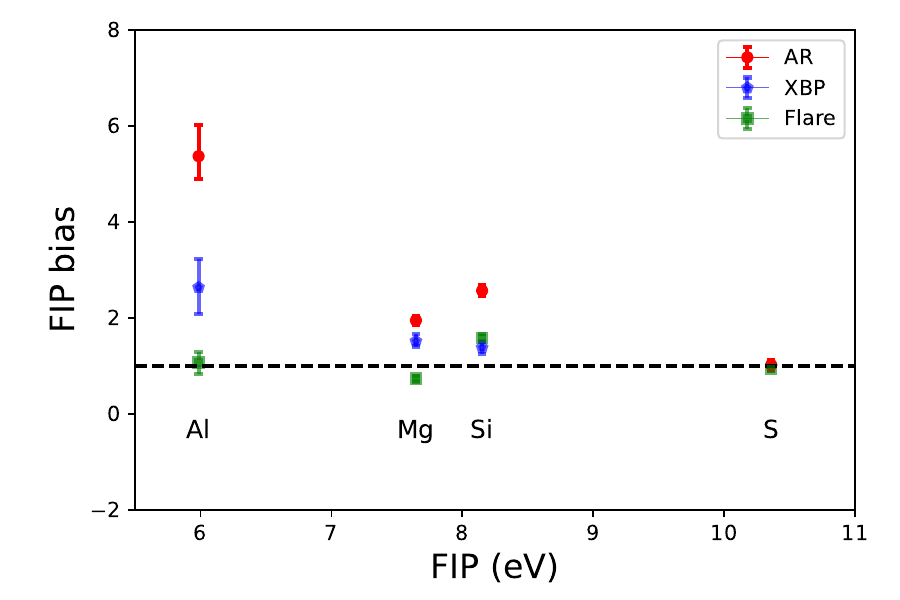}
    \caption{Variation of the FIP bias with the FIP of the elements. The red points are the averaged FIP bias for one of the AR (AR 12749) reported in the present study.
    The blue points are the FIP bias for the XBPs as reported by \cite{xsm_XBP_abundance_2021}.
    The green points are the measured FIP bias during the solar flares peak as reported by \cite{Mondal_2021ApJ}.
\label{fig:FIP_variation}}
\end{figure}

Figure~\ref{fig:FIP_variation} red points show the average values of the FIP bias (relative to the photospheric abundance~\cite{Asplund_2009}) for all the elements as a function of their FIP values.
{ 
The lower FIP element, Al (FIP = 5.99), is found to have the highest FIP bias of 5-6, whereas the low-FIP elements, Mg (FIP = 7.65) and Si (FIP = 8.15), are found to have a lower FIP bias of 2-2.5. The mid/high FIP element, S, is found to have a much lower FIP bias close to unity. 
A higher FIP bias for Al is noteworthy and may point to an intriguing physical process. However, this may also be a modeling artifact.}

One of the possibilities could be due to missing flux 
caused by the presence of multi-thermal plasma providing strong signals from  emission lines of Al or Mg formed at different temperatures.
To verify this we have simulated the emission lines in the energy range of the Mg/Al line complex by considering both the isothermal model and a multi-thermal model using the AR DEM of AR12759, reported by~\cite{Zanna_2022ApJ} (see Figure~\ref{fig-spectra_Al_lineComplex} in Appendix~\ref{appendix_a}). 
Similar line intensities from various ionization stages of Al and Mg can be seen in both the isothermal and multi-thermal models, confirming that the absence of the flux is not the result of multi-thermal plasma.

Another possibility is that missing flux is caused by missing lines of Al or Mg (mostly satellite lines) that are not yet present in CHIANTI version 10. 
We have analysed the high-resolution spectroscopic observations  described by \cite{Walker_1974ApJ...188..423W} and found several 
observed lines that are missing in the database. 
However, the total missing flux, compared to the predicted flux
by CHIANTI is not enough to explain the anomalous Al abundance.  
However, the \cite{Walker_1974ApJ...188..423W} 
observations were taken during a high level of solar activity, so it is possible that the missing lines have a 
stronger contribution at 3 MK. 
The Al abundance is currently clearly
overestimated by some degree.

Although this analysis is not conclusive enough to 
rule out Al's high FIP bias as an artifact, it is also not sufficient to conclude that it is not real. A higher Al FIP bias could be real.
This might be explained by examining a few particular scenarios from the Ponderomotive force model~\citep{Laming_2015} proposed by Laming (private communication), which could be investigated in a subsequent study.

\begin{deluxetable*}{c c c c c c c}[!ht]
\tablecaption{Best fitted parameters for the average spectrum of each AR.}
\label{table-1}
\tablehead{
AR&T & EM & Mg & Al & Si & S\\
&(MK) & (10$^{46}$ cm$^{-3}$) &  &  &   &
}
\startdata
12749&$ 3.22^{+0.04}_{-0.04} $&  $ 2.78^{+0.21}_{-0.20}$&  $7.93^{+0.02}_{-0.02}$&  $7.22^{+0.04}_{-0.04}$&  $7.90^{+0.02}_{-0.02}$&  $7.07^{+0.04}_{-0.05}$\\
12759&$ 3.24^{+0.04}_{-0.02} $&  $ 5.41^{+0.23}_{-0.35}$&  $7.86^{+0.02}_{-0.02}$&  $7.15^{+0.05}_{-0.04}$&  $7.94^{+0.01}_{-0.01}$&  $7.13^{+0.02}_{-0.03}$\\
12758&$ 3.04^{+0.05}_{-0.03} $&  $ 4.14^{+0.28}_{-0.36}$&  $7.87^{+0.03}_{-0.02}$&  $7.15^{+0.05}_{-0.05}$&  $7.92^{+0.02}_{-0.02}$&  $7.18^{+0.04}_{-0.06}$\\
%
%
\enddata
\end{deluxetable*}

{FIP bias of the quiet Sun \citep{xsm_XBP_abundance_2021} and during flares \citep{Mondal_2021ApJ} have been studied earlier using the XSM. For visual inspection, a combined plot is generated (Figure~\ref{fig:FIP_variation}). 
The blue points depict the FIP bias during the quiet period of the Sun when the disk emission is dominated by X-ray Bright Points (XBP: \citealp{xsm_XBP_abundance_2021}). On the other hand, the green points represent the FIP bias during the peak of the solar flares, as reported by \cite{Mondal_2021ApJ}. 
In our present study, the FIP bias of an AR core (red points) shows a consistently higher value for all the elements such as Al, Mg, and Si compared to that of the XBPs (green points). Since ARs possess a much higher magnetic field compared to XBPs, Ponderomotive force may play a crucial role~\citep{Laming_2015} in deciding the high value of FIP bias in ARs. However, due to chromospheric evaporation, FIP bias becomes near unity during flares. 
It is worth mentioning here that the abundance of Oxygen (8.89 dex) considered in all the above-mentioned earlier works was the value prescribed by \cite{Fludra_1999}. 
Eventhough in our present study we consider a very similar value (8.8 dex), for the confirmation's sake, we also have verified that the present results remain unchanged even if we change our Oxygen abundance value to the value of ~\cite{Fludra_1999}.}

\section{Summary} 
  \label{sec:summary}

We present the evolution of plasma characteristics for three ARs using disk-integrated soft X-ray spectroscopic observations from the XSM to make simultaneous line and continuum measurements.
Carrying out a comprehensive study of an AR using the Sun-as-a-star mode observation is challenging because of the presence of multiple activities throughout the solar cycle.
Unique XSM observations made during the minimum of Solar Cycle 24 allowed the study of the evolution of temperature, EM, and the abundances of Mg, Al, and Si for the individual ARs in the absence of any other noteworthy activity on the solar disk. 
Since the ARs were the principal contributors of disk-integrated X-rays during their evolution, the temperature and EM followed their X-ray light curve.
The average temperature of all the ARs is $\sim$3 MK, close to the well-known temperature of the AR core.
Irrespective of the activity and age of the ARs, the abundances or the FIP biases of Al, Mg, and Si were found to be consistently greater than their photospheric values, without much variation.
The abundance values develop within $\sim$10 hours of the appearance of the AR during its emerging phase. 
{ Throughout the AR evolution, the low FIP elements, Mg and Si, have a FIP bias of 2-2.5, whereas the mid-FIP element, S, has an average FIP bias of almost unity. 
The lowest FIP element, Al, has a greater FIP bias of $\sim$5-6.  
}
After discussing various modeling artifacts, the Al abundance appears to be overestimated, although the exact factor is unknown. Increased Al abundance could be real, implying that low-FIP elements degree of FIP bias is linked to their FIP values. 
Future spectroscopic studies to measure the FIP bias for more low-FIP elements (for example, Ca, whose FIP bias is between Al and Mg) would help us to better understand this phenomenon. In this regard, recent and upcoming X-ray spectrometers (for example,  DAXSS:~\citep{Schwab_2020ApJ} onboard INSPIRESat-1,
SoLEXS~\citep{Sankarasubramanian_2011ASInC} onboard upcoming Aditya-L1 observatory, and rocket-borne spectrometer MaGIXS~\citep{Champey_2022JAI}) will be useful. 

{ Finally, we stress that the abundances of Mg, Al, Si, and S are obtained from the line-to-continuum measurement in the 1.3 to 3 keV energy range, where the continuum is determined by the abundance of other elements, primarily oxygen. We assumed an average oxygen abundance obtained for solar wind, and we have justified this choice. However, as the solar wind measurements indicate a significant O variability, 
our results have some uncertainty in terms of absolute values (with respect to the H) of abundance and FIP bias. In the future, for better measurement of the abundance close to the absolute scale, a simultaneous measurement of the oxygen abundance is needed.}

\acknowledgments{
We acknowledge the use of data from the Solar X-ray Monitor (XSM) on board the Chandrayaan-2 mission of the Indian Space Research Organisation (ISRO), archived at the Indian Space Science Data Centre (ISSDC). 
The XSM was developed by the engineering team of Physical Research Laboratory (PRL) lead by Dr. M. Shanmugam,  with support from various ISRO centers.
We thank various facilities and the technical teams from all contributing institutes 
and Chandrayaan-2 project, mission operations, and ground segment teams for their support.
Research at PRL is supported by the Department of Space, Govt. of India. 
We acknowledge the support from Royal Society through the international exchanges grant No. IES{\textbackslash}R2{\textbackslash}170199.
GDZ and HEM acknowledge support from STFC (UK) via the consolidated grant to the atomic astrophysics group
at DAMTP, University of Cambridge (ST{\textbackslash}T000481{\textbackslash}1).
AB was the J C Bose National Fellow during the period of this work.
We thank Dr. Martin Laming for the useful discussion on anomalous Al abundance. We are thankful to an anonymous referee for providing us with very useful feedback.}


\appendix 
\renewcommand\thefigure{\thesection.\arabic{figure}}
\setcounter{figure}{0}

\section{Results of MCMC analysis}\label{app:MCMC_results}

We carried out Markov Chain Monte Carlo (MCMC) analysis of the spectra to obtain the regions of parameter space that best fits the observed spectra. This was done using the `chain' method available within XSPEC. Figure A1 shows the corner plot of the results for the spectrum on 01-Oct-2019. The results show that all parameters are well constrained by the spectra. Particularly, we note that there is no anti-correlation observed between Al and Mg abundances showing that the enhances Al abundances obtained cannot be adjusted by enhancements in Mg abundances. Similar trends are observed for spectra of other days as well. 



\begin{figure*}[!ht]
\begin{center}
    \includegraphics[width=0.8\textwidth]{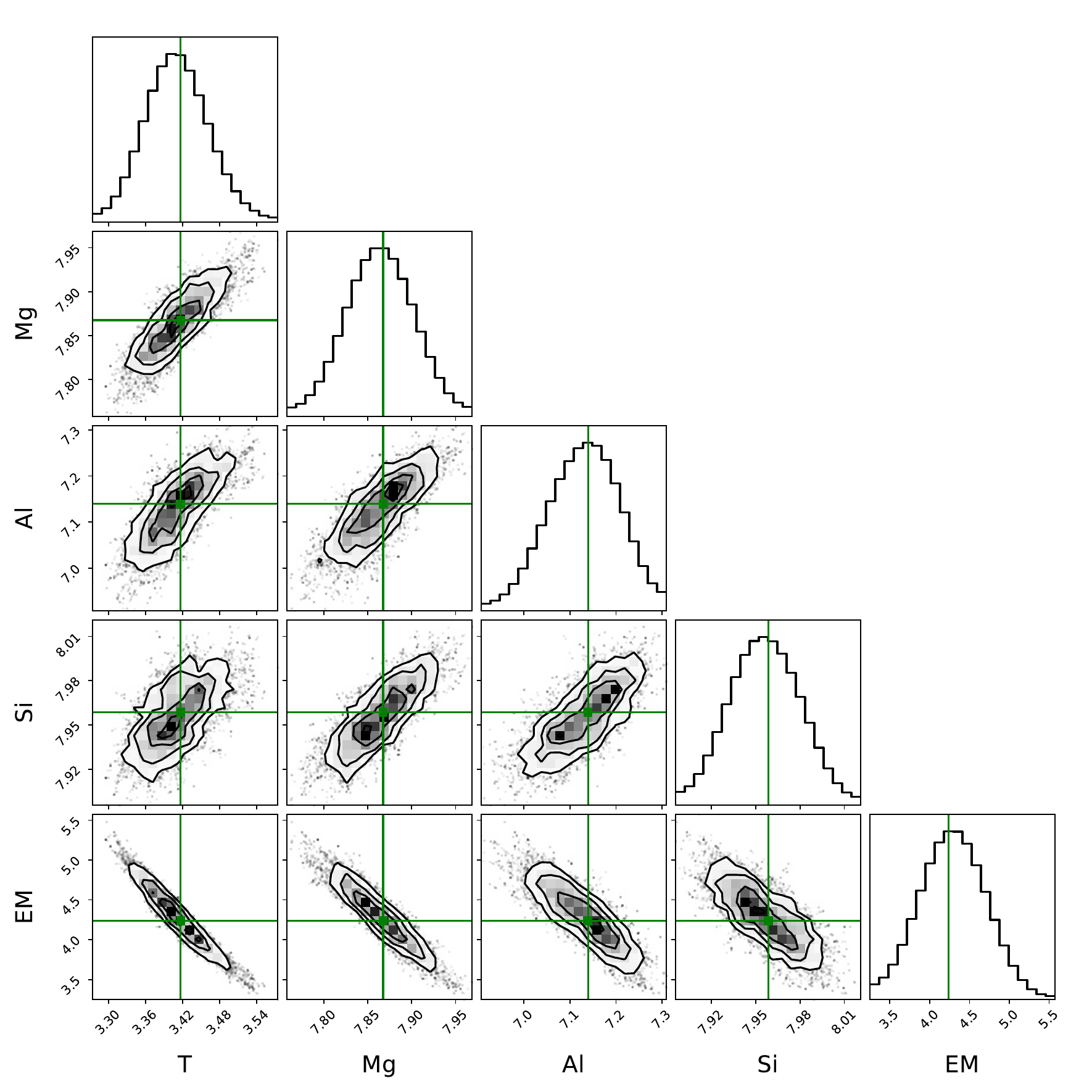}
    \caption{Corner plot depicting the results of MCMC analysis for the fitted spectrum on 01-Oct-2019. The histograms depict the marginalized distribution associated with each parameter. The scatter-plots are overlaid with contours representing 1$\sigma$, 2$\sigma$, and 3$\sigma$ levels to show correlations between all parameters. The best-fit parameters are represented by green lines. 
\label{fig:MCMC_results}}
\end{center}
\end{figure*}

\section{Simulated Spectrum}\label{appendix_a}

To check the effect of temperatures on the Mg/Al line fluxes in the XSM energy range of 1.55 to 1.70 keV, we have compared the simulated spectra in the same energy range by considering the isothermal and multi-thermal DEM models.
Figure~\ref{fig-spectra_Al_lineComplex} shows the simulated 3 MK spectrum (blue) overplotted with the  multithermal spectrum (red). 
The isothermal spectrum is generated for an emission measure of 10$^{27}$ cm$^{-5}$. 
The multithermal spectrum is derived by using the reported quiescent AR DEM by \cite{Zanna_2022ApJ}, which was obtained from the Hinode EIS observation of AR12759. 
For the comparison of both spectra, we have normalized them with the corresponding line flux of Mg XI, and Al XI-XII.
Similar line intensities predicted by both isothermal and multithermal models indicates that spectra are insensitive to temperature in this case.

\setcounter{figure}{0}

\begin{figure}
\centering
\includegraphics[width=1\linewidth]{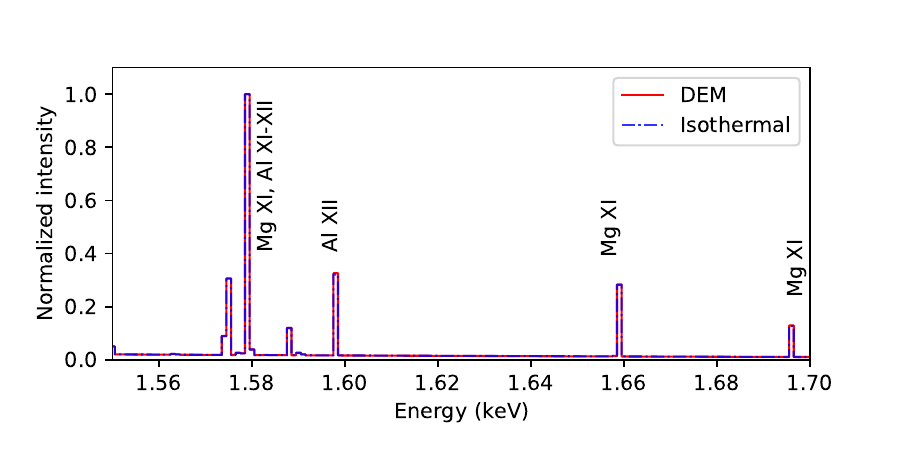}
\caption{Simulated spectra from CHIANTI v 10 in the energy range of Mg/Al line complex of XSM observed spectrum. Solid blue curve show the multi-thermal spectrum and dashed orange curve shows the isothermal spectrum.  
}
\label{fig-spectra_Al_lineComplex}
\end{figure}


\section{ Dependence of continuum on the abundant elements}\label{app_emission_procc}

\begin{figure}
\centering
\includegraphics[width=1\linewidth]{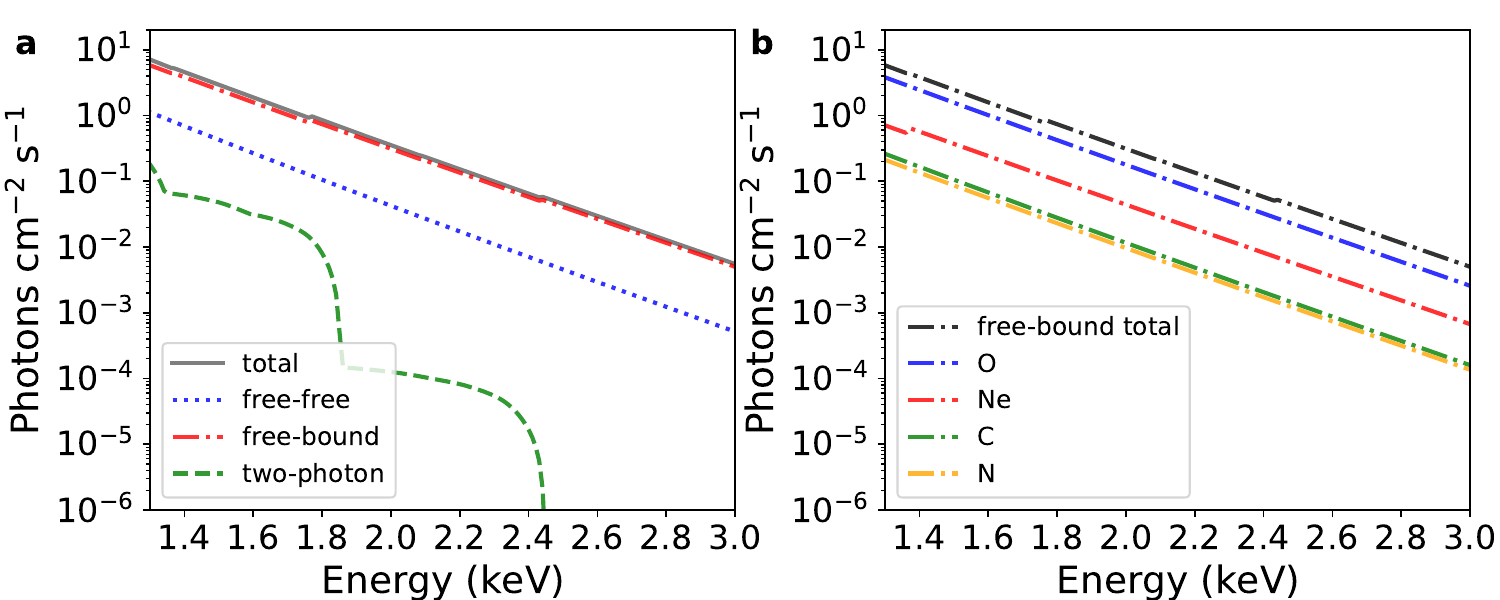}
\caption{ (a) Different component of continuum spectrum in the energy range of 1 keV to 3 keV at 3 MK. (b) Contribution of different elements on free-bound continuum. Total free-bound continuum is shown by black color.  
}
\label{fig-ap3}
\end{figure}

{ We have measured the abundances of Mg, Al, Si, and S with respect to $1$ to $3$ keV continuum emission produced by three physical processes -- thermal Bremsstrahlung, free-bound, and two-photon. Among these three processes at an AR temperature of around 3 MK, the contribution of two-photon is almost negligible ($<$2$\%$), whereas the free-bound emission contributes more than 80$\%$ (see Figure~\ref{fig-ap3}b).
Thus our measured abundances of Mg, Al, Si, and S will depend on the abundances of all other elements (except Mg, Al, Si, and S) that contribute to the free-bound and free-free emission. Hydrogen primarily determines the free-free continuum; however, the other heavy elements contribute to the free-bound emission. 
Among all of the elements, O and Ne are the primary contributors (nearly 80$\%$) to the free-bound continuum (see Figure~\ref{fig-ap3}b).}

\newpage

\bibliography{sola_bibliography_example}
\bibliographystyle{aasjournal}

\end{document}